\documentclass[12pt,preprint]{emulateapj}
\usepackage{graphicx,longtable}

\shortauthors{Dragomir et al.}
\slugcomment{Submitted for publication in the Astrophysical Journal}

\begin{document}

\renewcommand{\bottomfraction}{0.9}

\title{The HD 192263 system: planetary orbital period and stellar variability disentangled}

\author{
  Diana Dragomir\altaffilmark{1},
  Stephen R. Kane\altaffilmark{2},
  Gregory W. Henry\altaffilmark{3},
  David R. Ciardi\altaffilmark{2},
  Debra A. Fischer\altaffilmark{4},
  Andrew W. Howard\altaffilmark{5,6},
  Eric L. N. Jensen\altaffilmark{7},
  Gregory Laughlin\altaffilmark{8},
  Suvrath Mahadevan\altaffilmark{9,10},
  Jaymie M. Matthews\altaffilmark{1},
  Genady Pilyavsky\altaffilmark{9},
  Kaspar von Braun\altaffilmark{2},
  Sharon X. Wang\altaffilmark{9},
  Jason T. Wright\altaffilmark{9,10}
  }

\email{diana@phas.ubc.ca}
\altaffiltext{1}{Department of Physics \& Astronomy, University of
  British Columbia, Vancouver, BC V6T1Z1, Canada}
\altaffiltext{2}{NASA Exoplanet Science Institute, Caltech, MS 100-22,
  770 South Wilson Avenue, Pasadena, CA 91125}
\altaffiltext{3}{Center of Excellence in Information Systems, Tennessee
  State University, 3500 John A. Merritt Blvd., Box 9501, Nashville,
  TN 37209}
\altaffiltext{4}{Department of Astronomy, Yale University, New Haven,
  CT 06511}
\altaffiltext{5}{Department of Astronomy, University of California,
  Berkeley, CA 94720}
\altaffiltext{6}{Space Sciences Laboratory, University of California,
  Berkeley, CA 94720}
\altaffiltext{7}{Department of Physics $\&$ Astronomy, Swarthmore College, Swarthmore, PA 19081}
\altaffiltext{8}{UCO/Lick Observatory, University of California, Santa Cruz, CA 95064}
\altaffiltext{9}{Department of Astronomy and Astrophysics,
  Pennsylvania State University, 525 Davey Laboratory, University
  Park, PA 16802}
\altaffiltext{10}{Center for Exoplanets \& Habitable Worlds,
  Pennsylvania State University, 525 Davey Laboratory, University
  Park, PA 16802}


\begin{abstract}

As part of the Transit Ephemeris Refinement and Monitoring Survey
(TERMS), we present new radial velocities and photometry of the HD
192263 system. Our analysis of the already available Keck-HIRES and CORALIE radial
velocity measurements together with the five new Keck measurements we
report in this paper results in improved orbital parameters for the
system. We derive constraints on the size and phase location of the
transit window for HD 192263b, a Jupiter-mass planet with a period of
24.3587 $\pm$ 0.0022 days. We use 10 years of Automated Photoelectric Telescope (APT)
photometry to analyze the stellar variability and search for planetary
transits. We find continuing evidence of spot activity with periods near 23.4 days. 
The shape of the corresponding photometric variations changes over time,
giving rise to not one but several Fourier peaks near this value. However,
none of these frequencies coincides with the planet's orbital period and thus
we find no evidence of star-planet interactions in the system. We attribute
the $\sim$23-day variability to stellar rotation. There are also indications of
spot variations on longer (8 years) timescales. Finally, we use the photometric
data to exclude transits for a planet with the predicted radius of
1.09 $R_{J}$, and as small as 0.79 $R_{J}$. 
\end{abstract}



\section{Introduction}
\label{introduction}

Detecting exoplanets via the radial velocity (RV) method and subsequently monitoring their transit windows is the most fruitful strategy for exoplanets orbiting bright stars, which represent the best candidates for atmospheric studies. Ground and space-based transit surveys have revealed nearly 200 transiting exoplanets, but most of those orbit faint stars, while the RV technique is best suited for brighter stars. Moreover, searching for transits of known RV planets allows the selective monitoring of intermediate and long-period planets, of which only a few are known to transit so far. The Transit Ephemeris Refinement and Monitoring Survey (TERMS; \citealt{Kan09}) aims to improve the orbital parameters of RV exoplanets and monitor them photometrically during their thusly constrained transit windows.

In \cite{Dra11}, we presented new transits and improved parameters for several known transiting exoplanets, originally discovered by the SuperWASP survey \citep{Poll06}. We demonstrated that the photometric precision required to detect and characterize transits of giant planets is easily attainable by modest-sized, ground-based facilities such as the Cerro Tololo Inter-American Observatory (CTIO ) 1.0 m telescope. Through TERMS, the ephemerides of HD 156846b, HD 114762b, HD 63454b and HD 168443b have been refined and transit searches conducted in each case (see \citealt{Kan11a}, \citealt{Kan11b}, \citealt{Kan11c} and \citealt{Pil11}, respectively). 

Exoplanets discovered using the RV method can sometimes be controversial, and HD 192263b is one such mischief. It was first published in 2000 \citep{San00} as a planet with a period of 24.13 days and $m \sin i$ = 0.73 M$_{J}$. These results arose from an analysis of RV measurements obtained using the CORALIE spectrograph. A paper by \cite{Vog00} followed, reporting a similar solution based on their Keck measurements, but noting that the chromospheric activity appeared to vary with a period close to that of the suspected planet. Two years later, \cite{Hen02} attributed the RV signal at least partly to stellar variability, as indicated by their photometric and spectrophotometric data. Indeed, a modulation with a period of $\sim$24 days is clearly visible in the light curves, and the power spectrum of the Ca II H and K spectrophotometric observations exhibits a significant peak at the same period. In the end, unlike other planet-like RV signals that were caused by stellar activity (\citealt{Que01}; \citealt{Des04}; \citealt{Pau04}), HD 192263b made a convincing comeback in \cite{San03}. New CORALIE measurements demonstrated that the RV variation remained coherent in amplitude and phase for over three years, while new photometric observations from La Palma revealed significant changes over time \citep{San03}.

In this paper, we present new Keck RV observations which we use to refine the orbital parameters (section 2) and the transit ephemeris of the revived HD 192263b (section 3). We introduce new photometry of the host star obtained between 2002 and 2011 in section 4, and report on the stellar variability during this period in section 5. Finally, in section 6 we analyze the photometric measurements acquired during the predicted transit window and exclude transits with the predicted depth of 2.5$\%$ with high confidence. We conclude in section 7.


\section{RV Measurements and Revised Orbital Model}

\subsection{Stellar Properties}

A K2 dwarf, HD 192263 is a relatively bright star (V = 7.8) located at a distance of 19.3 parsecs \citep{Koe10}. It is also a BY Dra variable \citep{Kaz06}, a class of active stars that experience brightness variations due to their spotted surface.

We used Spectroscopy Made Easy (SME) on a spectrum of HD 192263 (taken without the iodine cell used for RV measurements), as detailed in \cite{Val09}. This procedure is based on that described in \cite{Val05}, with an added improvement for self-consistency between values of stellar properties obtained from spectroscopy and those determined from isochrones \citep{Val09}. The improved method derives a value for the stellar surface gravity (log $g$), effective temperature ($T_{eff}$), iron abundance ($[Fe/H]$) and alpha-element enrichment ($[\alpha/Fe]$) from the spectroscopic analysis. The last three parameters and the luminosity ($L$) are used to obtain a value for log $g$ from isochrone models (log $g_{iso}$). The spectroscopic analysis is run again, now with log $g$ fixed at the value of log $g_{iso}$ from the previous iteration. The loop continues until log $g$ and log $g_{iso}$ agree. For HD 192263, the effective temperature resulting from this method is $4996 \pm 44$ K, and the stellar radius (calculated from $L$ and $T_{eff}$) is 0.73 $\pm$ 0.01 R$_{\odot}$. \cite{Tak07} carried out a Bayesian analysis using the stellar parameters arising directly from the SME procedure (log $g_{iso}, T_{eff}$ and $[Fe/H]$) and a dense grid of theoretical evolutionary tracks, and found a stellar radius of 0.77 $\pm$ 0.02 R$_{\odot}$. This value agrees with our SME result within the 2$\sigma$ uncertainties. Finally, \cite{van09} estimated the radius of HD 192263 using spectral energy distribution (SED) fitting and obtained a value of 0.76 $\pm$ 0.02 R$_{\odot}$, which agrees with our SME result within the 1$\sigma$ uncertainties. We adopt the SME stellar radius value for our calculations but compute a predicted transit depth using the largest published value \citep{Tak07} as well, to show that transits can be excluded for both cases.

From the SME analysis we also find values for the effective temperature, surface gravity, iron abundance, projected rotational velocity, and mass of the host star. The stellar properties are listed in Table 1.

\begin{deluxetable}{ccc}
\tabletypesize{\small}
\tablewidth{8.5cm}
  \tablecaption{\label{stellar} Stellar Properties}
  \tablehead{
    \colhead{Parameter} &
    \colhead{Value} &
    \colhead{Reference}
  }
  \startdata
  $V$                       & 7.767  &  \cite{Koe10} \\
  $B-V$                     & 0.957  &  \cite{Koe10} \\
  Distance (pc)             & 19.3  &   \cite{Koe10} \\ 
  $T_\mathrm{eff}$ (K)      & $4996 \pm 44$ & This work  \\
  $\log g$                  &  $4.628 \pm 0.060$ &  This work \\
  $[$Fe$/$H$]$              & $0.054 \pm 0.030$ &  This work \\
  $v \sin i$ (km\,s$^{-1}$) & $ < 1.0 \pm 0.50$ &  This work \\
  $log R'_{HK}$                       & $-4.56$ & \cite{Hen02}  \\
  $M_\star$ ($M_\odot$)     & $0.807 \pm 0.015$ &  This work \\
  $R_\star$ ($R_\odot$)     & $ 0.73 \pm 0.01$ & This work \\
  $age$ (Gyr)     &     $2.09 \pm 2.54$  &  This work
  \enddata
\end{deluxetable}

\begin{deluxetable}{ccc}
  \tablewidth{8.5cm}
  \tablecaption{\label{rvs1} CORALIE Radial Velocities}
  \tablehead{
    \colhead{Date} &
    \colhead{Radial Velocity} &
    \colhead{Uncertainty}  \\
    \colhead{(BJD -- 2440000)} &
    \colhead{(m\,s$^{-1}$)} &
    \colhead{(m\,s$^{-1}$)} 
  }
 \startdata
11355.814795	&	-64	&	4 \\
11367.830021	&	10	&	12 \\
11375.760201	&	-10	&	13 \\
11381.750648	&	-69	&	7 \\ 
11384.720313	&	-81	&	6 \\
11390.738894	&	16	&	7 \\
\enddata
\tablecomments{Table 2 is presented in its entirety in the electronic edition
of the Astrophysical Journal.  A portion is shown here for guidance regarding
its form and content.}
\end{deluxetable}

\begin{deluxetable}{ccc}
  \tablewidth{8.5cm}
  \tablecaption{\label{rvs2} Keck Radial Velocities}
  \tablehead{
    \colhead{Date} &
    \colhead{Radial Velocity} &
    \colhead{Uncertainty}  \\
    \colhead{(BJD -- 2440000)} &
    \colhead{(m\,s$^{-1}$)} &
    \colhead{(m\,s$^{-1}$)} 
  }
 \startdata
10984.063121	 &  32.75  &  1.30 \\  
11011.914062	 &  -11.45   &  1.28 \\ 	
11050.879437	 &  26.96   &  1.38 \\ 	
11069.890238	 & -32.85  &  1.16 \\ 	
11312.086411	&  -56.44  &  1.23 \\ 
11313.112603	 & -49.31  &  1.30 \\ 
11342.057784	 &  16.00  & 1.71 \\
11342.983844	 &  27.90  &  2.17 \\ 
11367.915954	 &  39.38  &  1.77 \\ 
11409.933630	 &  -39.58  &  1.96 \\ 
11411.879482	 &  -13.05  &  1.58 \\ 
11438.769456	 &   18.80 &   1.60 \\ 
11439.828250  &   28.04  &  1.72 \\ 
11440.887793	 &   30.68  &  1.87 \\ 
11441.833444	 &   42.11   & 1.62 \\ 
11704.020541  &  -30.66  &  1.67 \\ 
11793.833579  &  -55.58  &  1.84 \\ 
12004.133959	 &   34.71  &   2.13 \\ 
12008.138984	 &  7.54   &  1.45 \\ 
12031.056206	 &  15.41  & 1.54 \\
12063.060897	 &  -46.80 &  1.72 \\ 
12094.907368	 &  -2.04  &  1.54 \\ 
12128.910768	 &  30.51  &  2.13 \\ 
12391.141713  &  47.72  &  1.87 \\ 
12536.755506	 &  31.62  &  1.53 \\ 
12778.109319	 &  10.06  &  1.36 \\  
12833.916425	 &  42.13  & 1.63 \\ 
12853.999751	 &  17.87  &  1.59 \\ 
13181.024857	 &  -37.53 &  1.33 \\
13239.861434	 &  -4.41  &  1.04 \\
13546.990720	 &  -45.96  &  1.05 \\ 
{\bf 13969.022212}	 &  {\bf-19.25}  & {\bf 0.83} \\
{\bf 14810.722283}	 &  {\bf 1.86}  &  {\bf 1.14} \\	
{\bf 15043.798793}	 &  {\bf-5.86}  &  {\bf 1.04} \\
{\bf 15412.021408}	 &  {\bf 37.22}  &  {\bf 0.85}  \\
{\bf 15782.862190}   &  {\bf 33.76}  &  {\bf 0.98}   
\enddata
\tablecomments{The values in bold are the five new RV measurements we report in this paper.}
\end{deluxetable}

\subsection{RV Measurements}

A total of 181 CORALIE (an echelle spectrograph on the 1.2m Swiss telescope at La Silla, Chile; \cite{Udr00}) RV measurements have been reported in \cite{San03}, of which those included in \cite{San00} are a subset. In addition there are 31 published Keck measurements (\cite{But06}; \cite{Vog00}). To these we add 5 new Keck observations, acquired between 2006 and 2011. All Keck observations were made with the HIRES echelle spectrograph \citep{Vog94} on the 10-m Keck I telescope. The HIRES instrument uses an iodine cell through which the starlight passes before reaching the slit. Minute Doppler shifts in the features of the stellar spectrum are accurately measured against a wavelength reference provided by the dense set of absorption lines in the iodine spectrum. The RVs were extracted following the procedure described in \cite{How09}.

An offset of 10672 m s$^{-1}$ (the median of the CORALIE RVs) was added to each of the CORALIE RVs to place them on the same scale as the Keck RVs. The complete set of CORALIE and Keck RV measurements are given in Tables 2 and 3, respectively. We emphasize that all CORALIE measurements and all Keck measurements except the 5 we report in this paper, have been previously published.

\subsection{Keplerian Model}

We fit the set of 217 available RVs using a single-planet Keplerian model based on the techniques described in \cite{How10} and the partially linearized, least-squares fitting method of \cite{Wri09}. Analyzing the data from the two telescopes together rather than separately provides smaller uncertainties on the period and mid-transit time. We allowed for an RV offset between the Keck and CORALIE measurements (4.10 $\pm$ 1.93 m s$^{-1}$). We also allowed for an offset between Keck measurements taken before and after JD = 2453237, due to a CCD and optics upgrade on that date (1.61 $\pm$ 5.84 m s$^{-1}$). The $rms$ of the RV residuals is 13.14 m s$^{-1}$ and the $\chi_{red}^{2}$ is 9.05, likely because HD 192263 is such an active star and its variability is not properly accounted for by the uncertainties on the RV measurements. To correct for this, we added a jitter term ($\sigma_{j}$) in quadrature to the measurement uncertainties ($\sigma_{RV}$): $w^{2} = \sigma_{RV}^{2} + \sigma_{j}^{2}$, and used $w$ instead of $\sigma_{RV}$ for the Keplerian fit \citep{Wri05}. We chose $\sigma_{j} = 10.15$ m s$^{-1}$ to satisfy the condition $\chi^{2} = 1$. Furthermore, this value is equivalent to the excess found by \cite{San03} in the residuals of their RV measurements, and is within the range of 10 - 30 m s$^{-1}$ predicted by \cite{Saa98} from log $R'_{HK} = -4.56$ \citep{Hen02}. This added measure also reduces the rms to 12.42 m s$^{-1}$.

The parameter uncertainties were determined from the sampling distribution of each parameter through a non-parametric bootstrap analysis \citep{Fre81}. 

\cite{San03} find a long term trend in their CORALIE data, to which they fit a line with a slope of $4.8 \pm 0.8$ m s$^{-1}$ yr$^{-1}$. They are unable to determine the source of this trend, which could be due to the presence of another companion, or a RV variation caused by stellar activity. We also include a linear velocity trend ($dv/dt$) in our models. The best-fit value for the offset between the CORALIE and Keck RVs is now 5.14 $\pm$ 2.18 m s$^{-1}$ while the offset between Keck measurements before and after JD = 2453237 becomes -17.67 $\pm$ 9.21 m s$^{-1}$. The best fit value for the linear velocity trend (when analyzing the Keck and CORALIE data combined) is $0.0070 \pm 0.0017$ m s$^{-1}$ day$^{-1}$ (or $2.56 \pm 0.62$ m s$^{-1}$ yr$^{-1}$). We note that including a trend does not significantly lower the $\chi^{2}$. When fitting only the CORALIE data we find a linear trend with a slope of $5.21 \pm 0.70$ m s$^{-1}$ yr$^{-1}$ which agrees with that obtained by \cite{San03}. However, fitting only the Keck data (which spans a longer time range encompassing that sampled by the CORALIE measurements) gives a slope of $-0.37 \pm 0.73$ m s$^{-1}$ yr$^{-1}$, a value consistent with the absence of a trend. Finally, we repeated the analysis for each of the three data sets, this time using the original measurement uncertainties (no jitter term). The results are statistically consistent with those obtained with the jitter term included. We conclude that additional measurements are necessary to firmly determine the source of the linear trend present in the CORALIE data. For the reasons described above, and because the uncertainties on the resulting values for the period and mid-transit time are smaller, we adopt the solution without the trend.

The parameters from our Keplerian fit (both with and without a linear trend as a free parameter) are given in Table 4, together with those reported by \cite{But06} for comparison. The folded data and adopted model are plotted in Figure 1. Figure 2 shows a zoomed-in view of the shaded area (top panel) and the residuals to the RVs within this area (bottom panel).

\begin{figure}[!t]
\begin{flushleft}
\includegraphics[scale=0.22]{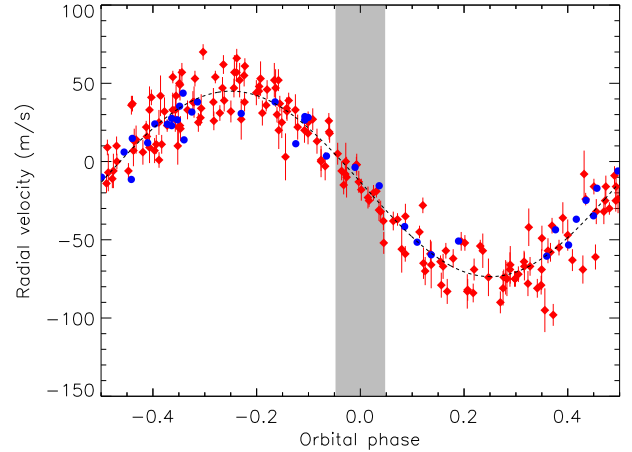}
\caption{CORALIE (red diamonds) and Keck-HIRES (blue dots) radial velocities. The error bars shown are the original measurement uncertainties, with the jitter term ($\sigma_{RV}$) not included. For most Keck measurements, the error bars are smaller than the size of the data points. The best-fit orbital solution is overplotted (dashed line). The shaded region corresponds to the 3$\sigma$ transit window and phase 0.0 is the predicted time of mid-transit. See section 2 in the text for details.}
\end{flushleft}
\end{figure}

\begin{figure}[!t]
\begin{flushleft}
\includegraphics[scale=0.22]{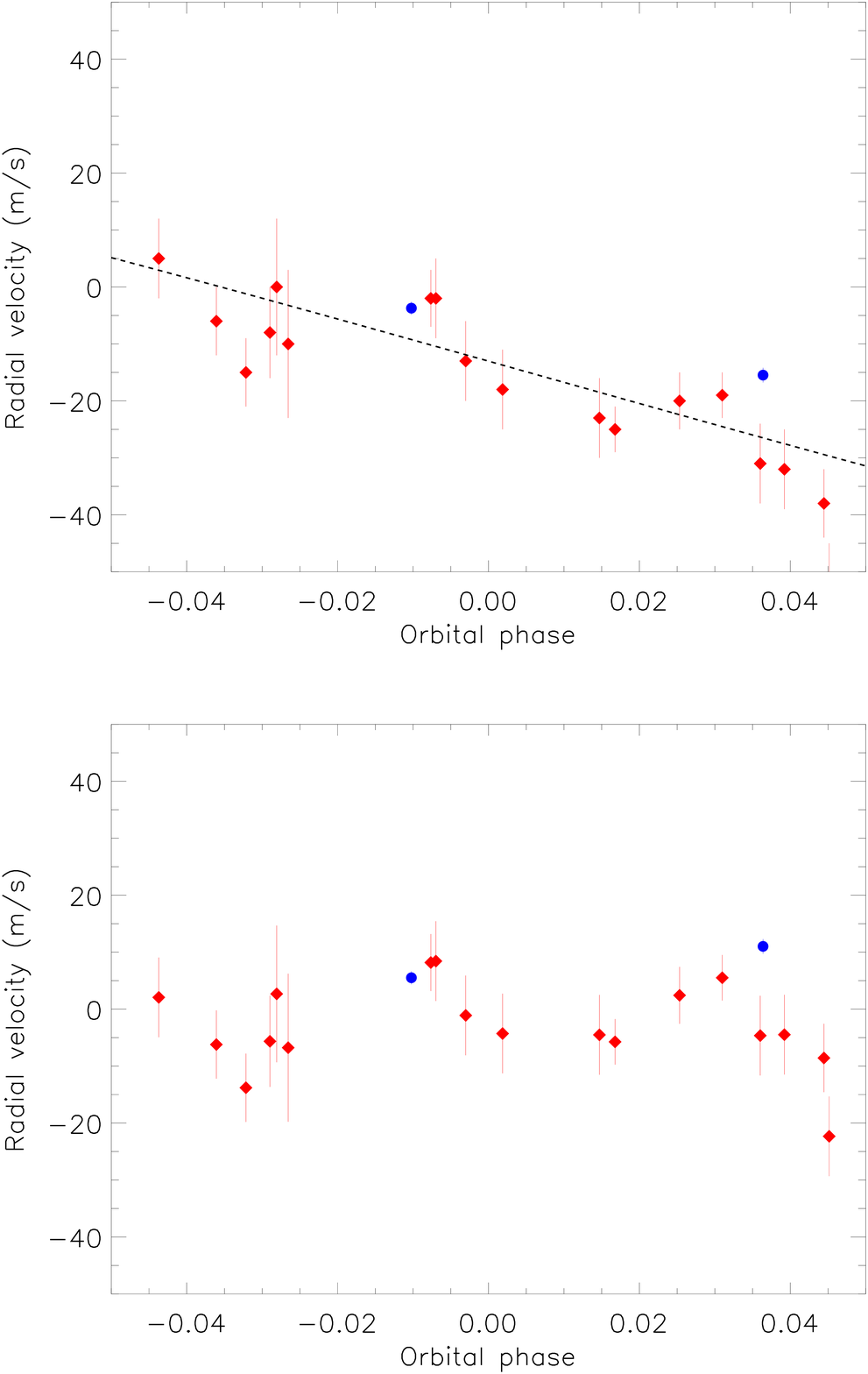}
\caption{Top panel: CORALIE (red diamonds) and Keck-HIRES (blue dots) radial velocities between the edges of the 3$\sigma$ transit window (shaded region in Figure 1), with the best-fit orbital solution overplotted (dashed line). Bottom panel: Residuals to the radial velocities, over the same phase range ($rms =$ 12.04 m s$^{-1}$). The error bars shown are the original measurement uncertainties, with the jitter term ($\sigma_{RV}$) not included.}
\end{flushleft}
\end{figure}

\begin{figure}[!t]
\begin{flushleft}
\includegraphics[scale=0.22]{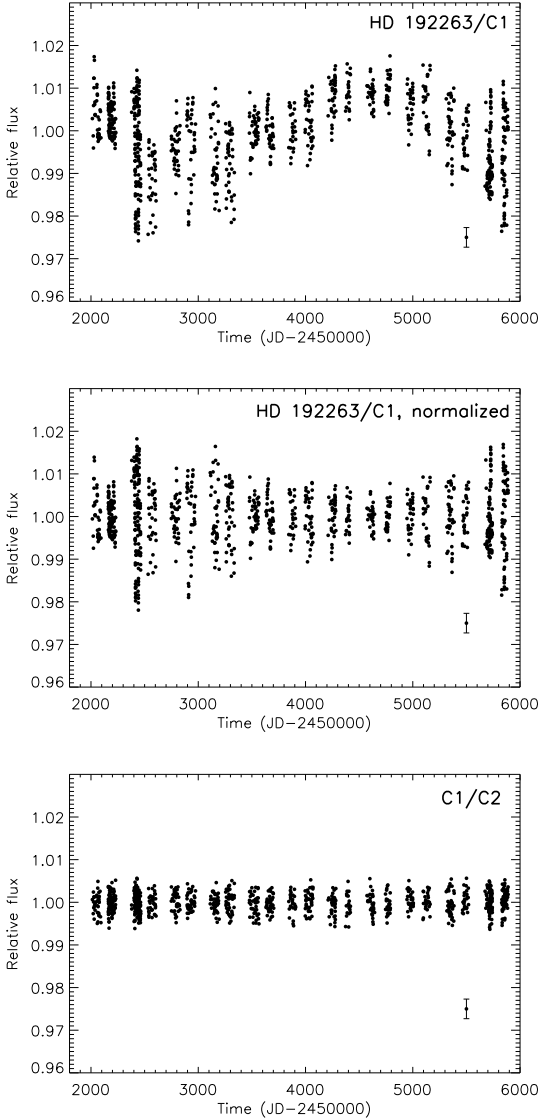}
\caption{T11 APT observations of HD 192263 covering 11 observing seasons from April 2001 to November 2011. The top panel: the photometry for HD 192263, obtained using comparison star C1 ($rms = 0.0085$). Middle panel: same as in the top panel, but with each half observing season normalized to the same mean ($rms = 0.0067$). Bottom panel: photometry of comparison star C1 relative to comparison star C2 ($rms = 0.0023$). It is clear that the large scatter in the HD 192263 observations is mainly due to the target star itself, not to any variability in the comparison stars. The 1$\sigma$ error bar is shown in each panel. See sections 4 and 5 in the text for details.}
\end{flushleft}
\end{figure}

\begin{deluxetable*}{lccc}
  \tablecaption{\label{planet} Keplerian Fit Parameters}
  \tablewidth{17cm}
  \tablehead{
    \colhead{Parameter} &
    \colhead{This work - without trend} & 
    \colhead{This work - with trend} &
    \colhead{Butler et al. 2006} \\
    \colhead{} &
    \colhead{(adopted)} &
    \colhead{} &
    \colhead{} 
  }
  \startdata
  $P$ (days)                       & $24.3587 \pm  0.0022 $ &  $24.3581 \pm 0.0028 $ & $ 24.3556 \pm 0.0046 $ \\
  $T_c\,^{a}$ (BJD -- 2440000)      & $10986.74 \pm 0.21$ & $  10986.73 \pm 0.21  $ & $ 10987.22 \pm 0.39 $ \\
  $T_p\,^{b}$ (BJD -- 2440000)      & $11796.9 \pm 6.8$  & $11795.0 \pm 4.6$ & $11994.3 \pm 3.9 $ \\
  $e$                              & $0.008 \pm 0.014$   & $ 0.008 \pm 0.014  $  & $ 0.055 \pm 0.039$ \\
  $K$ (m\,s$^{-1}$)                & $59.3 \pm 1.2$ &  $ 58.7 \pm 1.2  $ & $ 51.9 \pm 2.6 $ \\
  $\omega$ (deg)                   & $184 \pm 90$ & $ 157 \pm 88  $  & $ 200 $ \\
  $dv/dt$ (m\,s$^{-1}$\,yr$^{-1}$) & $ - $  & $  2.56 \pm 0.62  $  & $ - $ \\
  $M_p \sin i$ ($M_J$)             & $0.733 \pm 0.015$ & $ 0.726 \pm 0.014   $ & $ 0.641 $ \\
  $a$ (AU)                         & $0.15312 \pm 0.00095$  &  $ 0.15312 \pm 0.00095   $ & $ 0.15 $ \\
  rms (m\,s$^{-1}$)                &  12.42     & 11.87          & 12.5
  \enddata
  \tablenotetext{a}{Time of transit.}
  \tablenotetext{b}{Time of periastron passage.}
\end{deluxetable*}


\section{Predicted Transit Window and Characteristics}

From our newly derived stellar and planetary parameters we can ascertain the properties of the predicted transit. Using the models of \cite{bod03} and our revised measurement of the mass, we estimate a radius for the planet of R$_{p}$ = 1.09 R$_{J}$. It was shown by \cite{kan08} that the transit probability is a strong function of eccentricity and periastron argument. The eccentricity of this orbit is small enough such that this introduces a minor effect. Based upon our parameters (listed in the first column of Table 4), the transit probability is 2.49$\%$, the predicted duration is 0.192 days, and a predicted depth is 2.53$\%$, easily within the range of our photometric precision (rms = 0.0065 within the 3$\sigma$ transit window; see section 6 for details). 

We derived predicted transit times using the non-parametric bootstrap analysis described in Section 2.3. The predicted transit mid-point time used to fold the data was 2455882.84 $\pm$ 0.37 (JD). Note that the final Keck measurement obtained reduced the uncertainty on the transit mid-point from 0.47 days to 0.37 days, emphasizing the importance of extending the time baseline of the RV measurements when attempting to reduce the total size of the transit window. In this case, the 1$\sigma$ transit window has a total duration of 0.932 days. Although this is dominated by the transit mid-point uncertainty, it is small enough that we are able to attempt the detection of the transit.

As mentioned in Section 2.1, \cite{Tak07} derive a slightly larger stellar radius of 0.77 $\pm$ 0.02 R$_{\odot}$. Using this radius results in a transit probability of 2.61$\%$ and a transit depth of 2.28$\%$. This is still well within our photometric precision such that we are able to make a definitive statement regarding the transit exclusion of this planet.


\section{Photometry}

\subsection{APT photometry}

We acquired 985 photometric observations using the T11 0.8 m automatic photometric telescope (APT), located at Fairborn Observatory in southern Arizona. The data span just over a decade, from April 13, 2001 to November 23, 2011. Measurements were obtained simultaneously in the Str\"{o}mgren {\it b} and {\it y} passbands by two EMI 9124QB photomultiplier tubes. The individual {\it b} and {\it y} differential magnitudes are averaged to obtain the quantity $\Delta(b + y)/2$. The observing and data reduction procedures are identical to those described in \cite{Hen99} for the T8 0.8 m APT.

The differential magnitudes $\Delta(b+y)/2$ were converted to relative fluxes and normalized to 1 for the plots in Figure 3. Two comparison stars were considered for the differential photometry. Comparison star 1 (C1) is HD 193328 (V = 7.48, B-V = 0.12, A2), while comparison star 2 (C2) is HD 193225 (V = 7.35, B-V = 0.29, F0). Typical precision of a single relative flux measurement from T11 is $0.0010 - 0.0020$, depending on the quality of the night and the air mass of the observations. The standard deviation of the C1/C2 relative fluxes (see Figure 3, bottom panel) is slightly larger than this (0.0023). This is probably a combination of the air mass, since HD 192263 lies near the celestial equator, and also suspected low-amplitude variability in C2. For this reason, we analyzed the differential photometry of HD 192263 with respect to C1 (see Figure 3, top panel). The 985 $\Delta(b + y)/2$ magnitudes for HD 192263-C1 and C1-C2 are listed in Table 5.

HD 192263 is observable from Arizona only during part of the year, which explains the larger gaps in the data between observing seasons. In addition, the star is at opposition around July 24, during the annual Summer Shutdown of the APTs. This gives rise to the shorter (8 - 10 weeks) gaps between the first and second clusters of data points (half seasons), between the third and fourth, etc. As such, the first two sections of the light curve correspond to the first observing season, the third and fourth belong to the second observing season, and so on.

We prepared the data for the transit search by normalizing it such that each half season has the same mean value (see Figure 3, middle panel). This normalization affects the data on a timescale of months, thus preserving the shape and depth of any potential transits which would last less than 8 hours (see section 3).

\subsection{ASAS photometry}

HD 192263 was also observed as part of the All Sky Automated Survey\footnote{http://www.astrouw.edu.pl/asas/?page=aasc$\&$catsrc=asas3} (ASAS) \citep{Poj97}. A total of 345 data points are available in the ASAS-3 photometric V-band catalogue. These observations were collected between March 27, 2001 and November 10, 2009. The quality and cadence of this data set are lower than for the APT photometry. As a consequence, we use the ASAS observations to test our conclusions regarding the variability of the star, but we do not employ them for the transit search.

\begin{deluxetable}{ccc}
\tablewidth{8.5cm}
\tablecaption{Photometric Observations of HD 192263 from the T11 APT}
\tablehead{
\colhead{Heliocentric Julian Date} & \colhead{(HD 192263$-C1)_{by}$} & 
\colhead{$(C1-C2)_{by}$} \\
\colhead{(HJD $-$ 2,400,000)} & \colhead{(mag)} & \colhead{(mag)}
}
\startdata
52,012.9927 & 0.5293 & 0.0729 \\
52,020.9783 & 0.5337 & 0.0765 \\
52,022.9700 & 0.5368 & 0.0710 \\
52,023.9680 & 0.5260 & 0.0735 \\
52,025.9669 & 0.5191 & 0.0750 \\
52,027.9535 & 0.5215 & 0.0723 
\enddata
\tablecomments{Table 5 is presented in its entirety in the electronic edition
of the Astrophysical Journal.  A portion is shown here for guidance regarding
its form and content.}
\end{deluxetable}


\section{Variability of the host star}

In this section, we describe an investigation of the stellar variability of HD 192263 based on the 11 seasons of APT photometry described in the previous section. We generate an amplitude spectrum\footnote{An amplitude spectrum shows the amplitude of signals present in the data plotted versus frequency. A power spectrum is obtained by plotting the square of the amplitude versus frequency.} from a discrete Fourier transform of the time series (see top panel of Figure 4) and search it for significant peaks. Beyond the frequency range shown in Figure 4 (0.00 to 0.07 cycles day$^{-1}$), only harmonics of the significant frequencies shown in the top plot rise above the noise.

The middle panel and the inset of Figure 4 show the spectral window functions for the two significant frequencies shown in the top panel. These indicate the aliases introduced into the amplitude spectrum by the observing cadence, with regular gaps on nightly, biannual and annual timescales.

\begin{figure}[!t]
\includegraphics[scale=0.22]{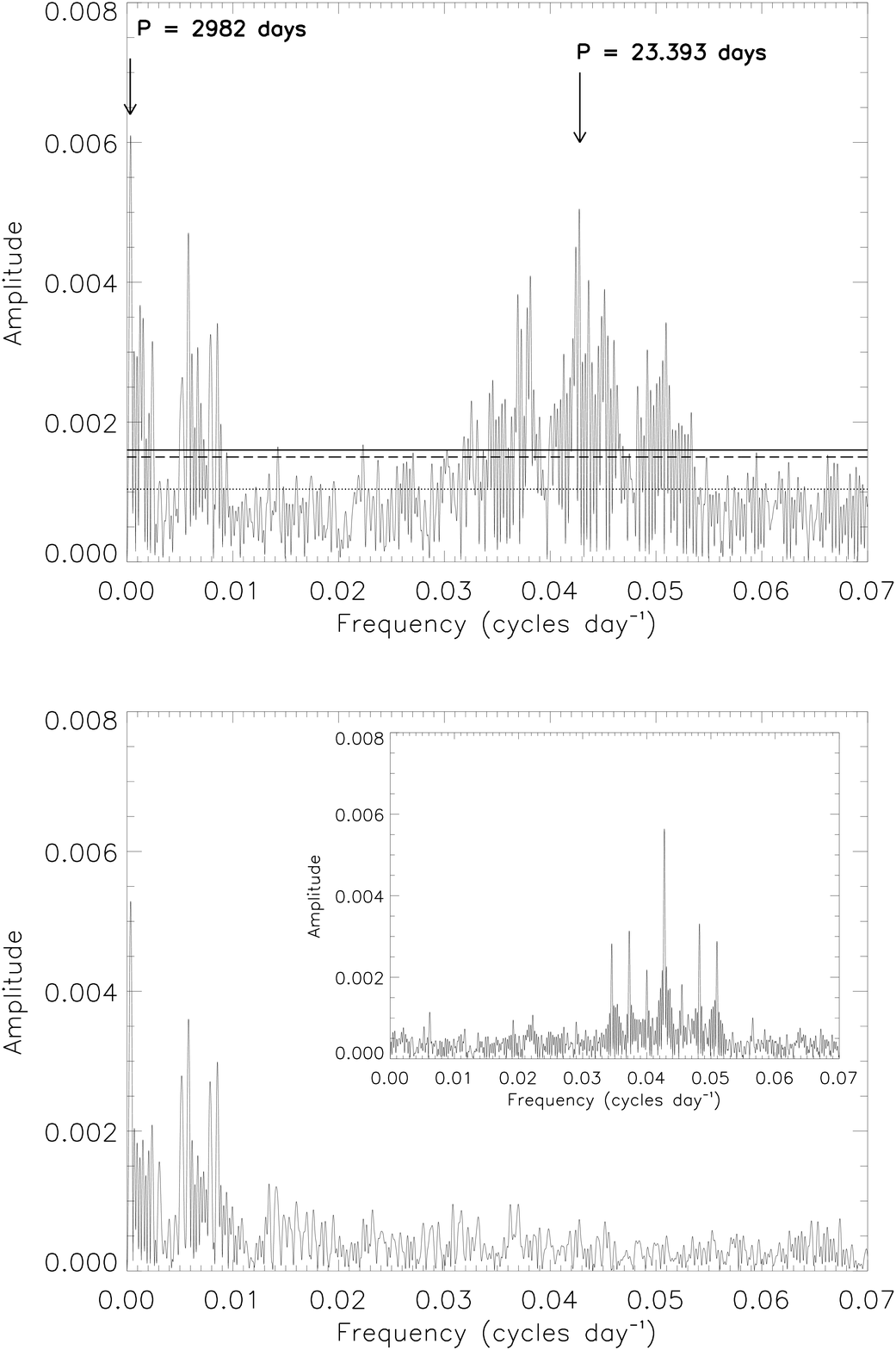}
\begin{center}
\includegraphics[scale=0.145]{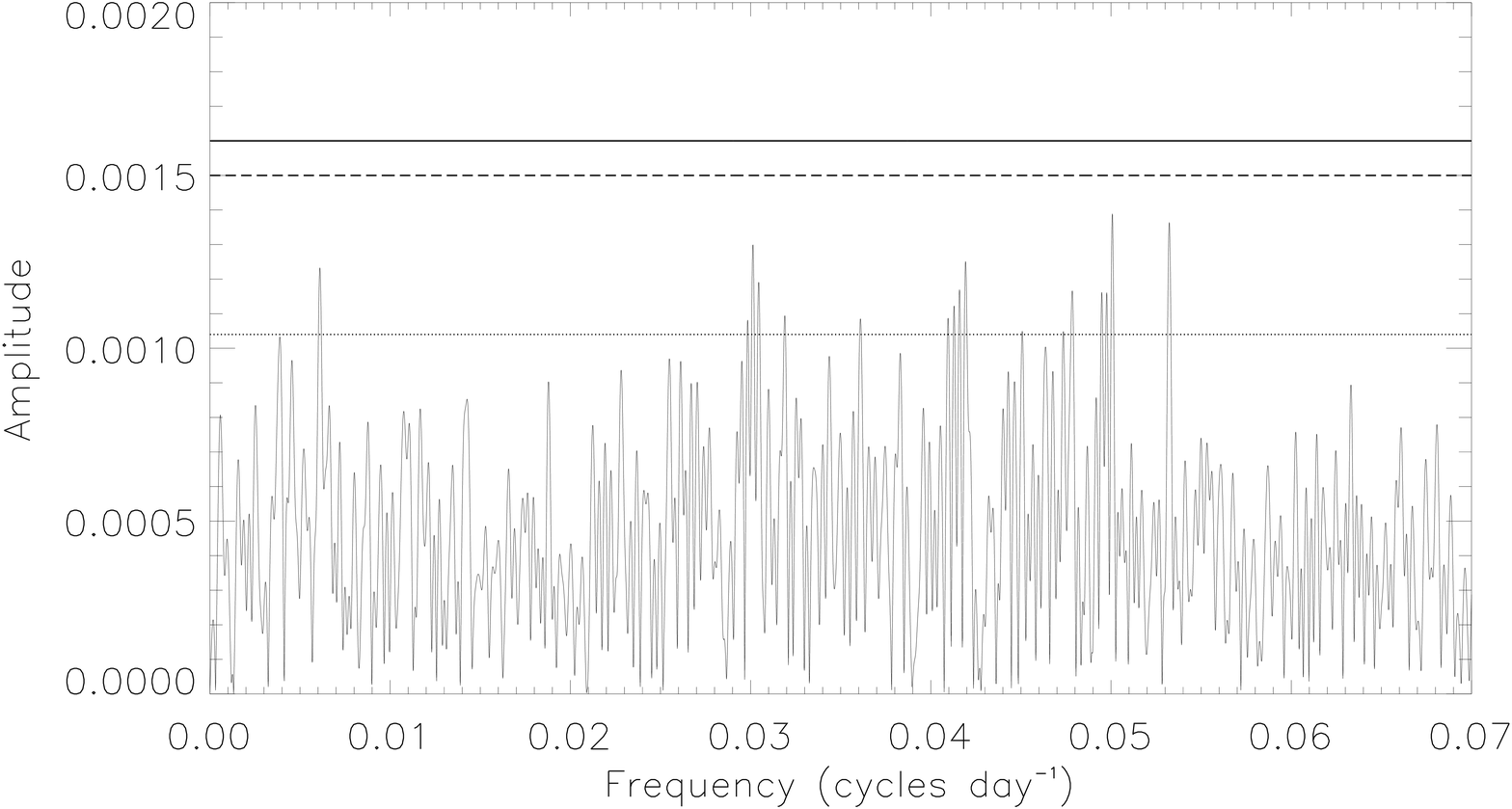}
\end{center}
\caption{Fourier amplitude spectrum (top), window functions for the two significant frequencies described in the text (middle and inset) and Fourier spectrum of the light curve residuals after pre-whitening (bottom) of the APT photometry. The dotted, dashed and solid horizontal lines in the top and bottom panels correspond to false-alarm probabilities of 4.6\%, 0.3\% and 0.1\%, respectively, calculated using the SigSpec routine \citep{Ree07}. Details can be found in section 5.}
\end{figure}

The strongest peak in the top panel of Figure 4 has a frequency of 0.0003354 $\pm$ 0.0000055 cycles day$^{-1}$, which is equivalent to a period of 8.17 years (or 2982 days $\pm$ 50). This signal is clearly visible in the photometry, as shown in the top panel of Figure 3. In light of the already known BY Dra nature of the star, this 8.2-year timescale is consistent with an activity cycle in an early K dwarf like HD 192263, but its cyclical or periodic nature can only be determined with additional observations. The brightness changes are the result of changes in the filling factor of spots or active regions on the stellar photosphere \citep{Ozd10}. They may be driven by a stellar dynamo \citep{Str05}, or may arise from random variations in a large number of spots on the surface of an active star without the requirement of a driving mechanism, as shown by the simulations of \cite{Eat96}.

The amplitude spectrum further shows a cluster of peaks between 0.03 and 0.06 cycles day$^{-1}$, of which the largest occurs at 0.0427475 $\pm$ 0.0000084 cycles day$^{-1}$ (period = 23.3932 $\pm$ 0.0046 days) with an amplitude of 0.005 mag. The window function for a sinusoid of this frequency and amplitude (see inset of Figure 4) reveals that not all of these peaks are aliases, but that there are other independent frequencies present in this range. We first "pre-whitened" the light curve by sequentially removing the frequencies of the largest peaks (and their harmonics), until we reached the noise level. Six frequencies in this range were identified; they are listed in Table 6, along with the very low frequency discussed above. We then performed simultaneous least-squares fits to the data, where the frequencies, amplitudes and phases of these peaks were allowed to float, with the values obtained from the amplitude spectrum as initial guesses. The Fourier amplitude spectrum of the light curve residuals following the pre-whitening is plotted in the bottom panel of Figure 4. None of the identified frequencies corresponds to the orbital frequency of the exoplanet HD 192263 b. In fact, when we include that frequency and allow the least-squares fit to iterate, the frequency migrates to one of those listed in Table 6. Despite the fact that there are at least six frequencies in a narrow range, it is clear that the orbital frequency of the planet is not one of them.


The Fourier peaks clustered around a period of about 23.4 days are a result of the evolving nature of star spots. As spots or groups of spots form, evolve and eventually disappear, contemporaneously or successively, the shape and amplitude of the light curve change over the course of one to a few stellar rotation periods. Thus, although the true stellar rotation period is close to 23.4 days, it cannot be determined to the frequency resolution of the time series.

The frequency analysis and the behaviour of the light curve are consistent with rotational modulation of the light output of an active K0 dwarf observed for about a decade. If the star is undergoing an activity cycle, then the spottedness will also change on that timescale (varying the amplitude of rotational modulation). In the light curve, shown in the upper two panels of Figure 3, one can see the changing width of the envelope of points over the 11-year time span. If the variation at a period around 23.4 days were constant in amplitude, that envelope would maintain the same width (since the photometric scatter of the comparison star remains uniform throughout the data set, as can be seen in the bottom panel of Figure 3). All indications are that the amplitude of this signal changes over the span of the $\sim$8-year variation, and not in a random way.


\begin{deluxetable}{cccc}
\tabletypesize{\small}
\tablenum{6}
\tablewidth{0pt}
\tablecaption{List of observed frequencies for HD 192263}
\tablehead{
\colhead{$\#$} & \colhead{Frequency (c/d)} &  \colhead{Period (days)} & \colhead{Amplitude (rel. flux)}\\
\colhead{} & \colhead{$\sigma$ frequency} & \colhead{$\sigma$ period} & \colhead{$\sigma$ amplitude} 
}
\startdata
1 & 0.0003354 & 2982 & 0.00612 \\
  &  0.0000055 & 50 & 0.00059 \\
2 & 0.037823 & 26.439 & 0.00196 \\
  &  0.000024 & 0.017 & 0.00034 \\
3 &  0.040741 & 24.5640 & 0.00273 \\
   &  0.000017 & 0.0085 & 0.00042 \\
4 & 0.0424 & 23.5849 & 0.0024 \\
 &  0.0015 & 0.8650 & 0.0012 \\
5 &  0.0427475 & 23.3932 & 0.00278 \\
 &  0.0000084 & 0.0046 & 0.00012 \\
 6 &  0.043625 & 22.9226 & 0.00333 \\
  & 0.000017 & 0.0090 & 0.00040 \\
7 & 0.04485 & 22.30 & 0.00256 \\
 &  0.00029 & 0.15 & 0.00095
\enddata
\tablecomments{The full least-squares fit solution including phases for these frequencies is available upon request.}
\end{deluxetable}

Our findings indicate that the periodic variability reported by Henry et al. (2002) and Santos et al. (2003) persists. To further verify this, we subtracted the 8.17-year signal from the data and phased the residuals at a period of 23.39 days (the largest amplitude signal in this period range), as shown in the top panel of Figure 5. A dominant variation, which remains roughly in phase throughout the time span of the photometry, stands out in this phase diagram.

We also see periodicity around 12 days in some of the half-season data sets, and we see the first harmonics of the periods near 23.4 days in the Fourier spectrum. This is in agreement with the period of 12.2 $\pm$ 0.1 days obtained by Henry et al. (2002) from their spectrophotometric Ca II H and K observations. They proposed that the 12.2-day period is half the stellar rotation period, arising in the data when active regions are present on opposite hemispheres of the star.

If we take a value of 23.4 days as the rotation period of the star and our stellar radius of 0.73 R$_{\odot}$, we find a rotational velocity of 1.58 km/s, which is consistent with our measured value of $v \sin i < 1.0 \pm 0.5$ km s$^{-1}$ (i.e. $v_{rot} \geq v \sin i$).

Using photometry spanning just over a year, Henry et al. (2002) found a rotation period of 24.5 $\pm$ 0.5 days, which agrees with the period of the planet within the uncertainties. This led Santos et al. (2003) to suggest the possibility of star-exoplanet interactions in the system. However, none of our new, more precise values of the dominant frequencies associated with rotation, based on a much longer data set, match the orbital period of the planet. Our least-squares test, trying to force a fit including the orbital period of the planet (24.3587 days), show that it is not part of a valid solution for the data. As an additional reality check, we phased the photometry at this period and found that the coherent signal visible in the top panel of Figure 5 disappears, as can be seen in the bottom panel of the same figure. Based on our data, we conclude that there is no indication of star-planet interactions in the HD 192263 system.

\begin{figure}[!t]
\begin{flushleft}
\includegraphics[scale=0.22]{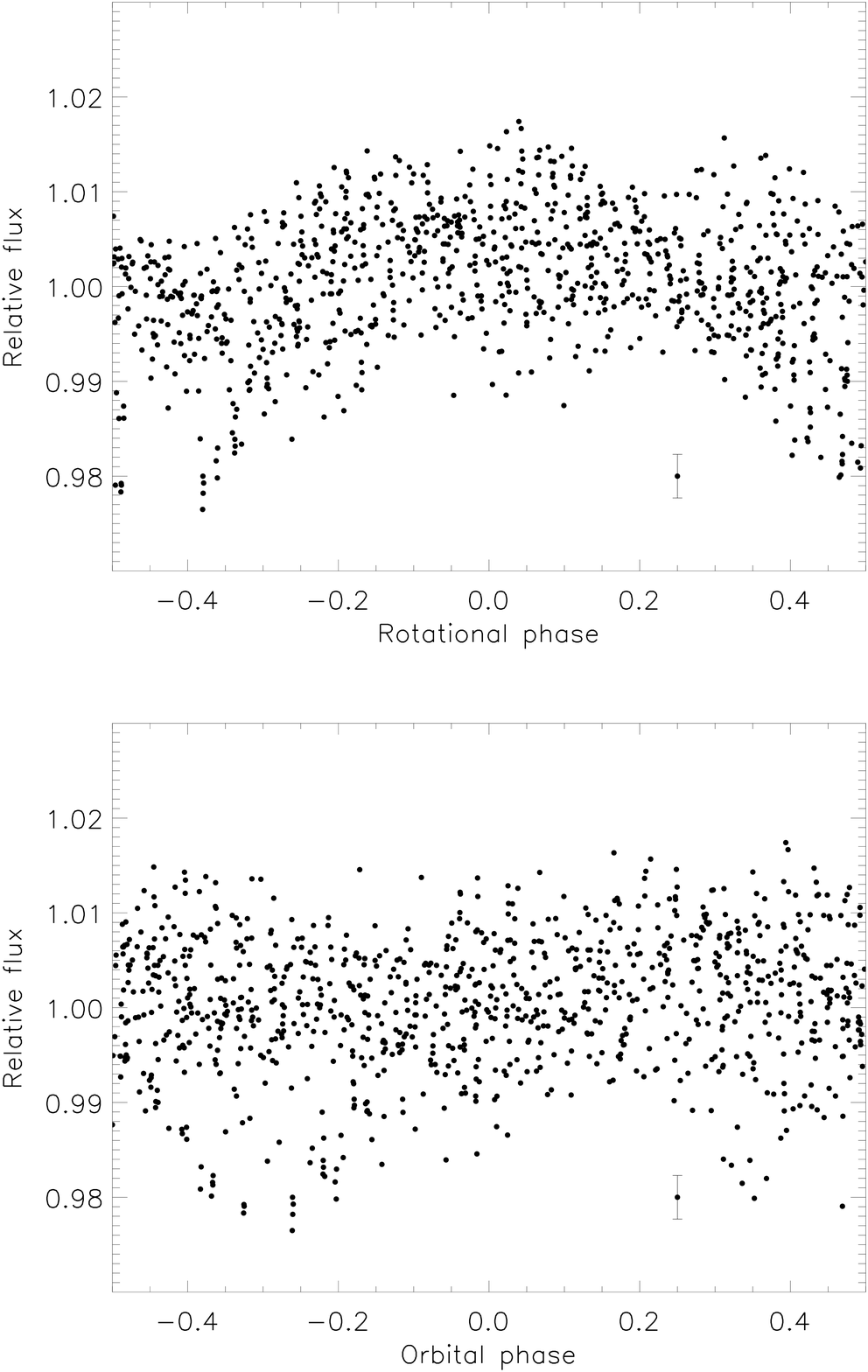}
\caption{APT photometry with 8.17-year trend removed, phased at the rotation period of the star (23.39 days; top panel) and at the orbital period of the planet (24.36 days; bottom panel). The coherent signal visible in the top panel disappears in the bottom panel, indicating that it is more likely to correspond to the stellar rotation period than to star-planet interaction. The 1$\sigma$ error bar is shown in each panel. Details can be found in section 5.}
\end{flushleft}
\end{figure}

For completeness, we have also examined the ASAS photometry with the goal of verifying the conclusions described above. This time series has a lower cadence, covers a shorter time frame and its rms is three times larger than for the APT data. We find that while the two data sets are consistent with each other, the lower quality of the ASAS photometry prevents it from setting additional constraints on the variability of the star.


\section{Transit Exclusion and Implications}

The long APT photometric dataset for HD 192263 makes possible the full coverage of the 3$\sigma$ transit window. Figure 6 shows the photometry phased at the planet's orbital period. Phase 0.0 is the location of the predicted mid-transit time of the planet. The solid line represents the predicted transit signature, based on the analytic models of \cite{Man02} and the values for the transit depth and duration calculated in section 3. The short dashed line corresponds to the predicted transit depth if the stellar radius is 0.77 R$_{\odot}$ (see section 3). At the top the entire orbital phase is shown, while the horizontal range of the bottom plot corresponds to the size of the 3$\sigma$ window. In the bottom panel, the vertical dashed and dotted lines indicate the extent of the 1$\sigma$ and 2$\sigma$ transit window, respectively. 

The rms of the photometry in Figure 6 is 0.0065. We can therefore exclude edge-on ($i$ = 90$^{\circ}$) transits with the predicted depth of 0.0253 at the 3.9$\sigma$ level. (If we assume a stellar radius of 0.77 R$_{\odot}$, then the predicted depth is 0.0228 and we can exclude such transits at the 3.5$\sigma$ level.) While a grazing transit for this planet requires an orbital inclination of 88.7$^{\circ}$, the cadence of the observations only allows us to exclude transits corresponding to $i > $88.9$^{\circ}$. This is equivalent to an impact parameter ($b$) $<$ 0.86, and a transit duration $>$ 0.01 days = 0.004 orbital phase. The photometric precision is sufficient to rule out transiting planets with radii as small as 0.79 $R_{J}$ at 2$\sigma$ confidence. The density of a planet with such a radius would fall outside the range predicted by the models of \cite{bod03}.

The case of HD 192263b shows that deep and more importantly, long transits can be detected or ruled out using existing low-cadence photometry if the dataset covers a sufficiently long time period. An additional benefit of a long time series is the full coverage of the 3$\sigma$ transit window, providing a higher confidence in the transit exclusion in the case of a non-detection. The utility of the ephemeris refinement component of the TERMS approach is clear: smaller uncertainties on the orbital period and transit time lead to a shorter transit window within which we need to assess whether the phase-folded photometry has adequate cadence and precision.

We perform an additional test to check the presence or absence of a transit by comparing the predicted amplitude of the Rossiter-McLaughlin (RM) effect for this system with our RV measurements. For a system in which the orbital plane of the planet is aligned with the equator of the star, \cite{Gau07} show that the amplitude of the RM effect is given by 

\begin{eqnarray}
K_{R}=v \sin i \frac{\gamma^{2}}{1-\gamma^{2}}
\end{eqnarray}

where $\gamma = R_{p}/R_{\star}$. For $\gamma \ll 1$, equation 1 becomes

\begin{eqnarray}
K_{R}=v \sin i  \left(\frac{R_{p}}{R_{\star}}\right)^{2}
\end{eqnarray}

Using a transit depth ($R_{p}/R_{\star})^2$ of 0.0253 and our upper limit on $v \sin i$, we find $K_{R} < 25.3$ m s$^{-1}$. Given the rms of our RV residuals within the 3$\sigma$ transit window, we can exclude an RM effect with amplitude $> 12.0$ m s$^{-1}$. While these limits place a much looser constraint on the existence of a transit than the photometry does, they are consistent with the absence of a transit.

\begin{figure}[!t]
\begin{flushleft}
\includegraphics[scale=0.22]{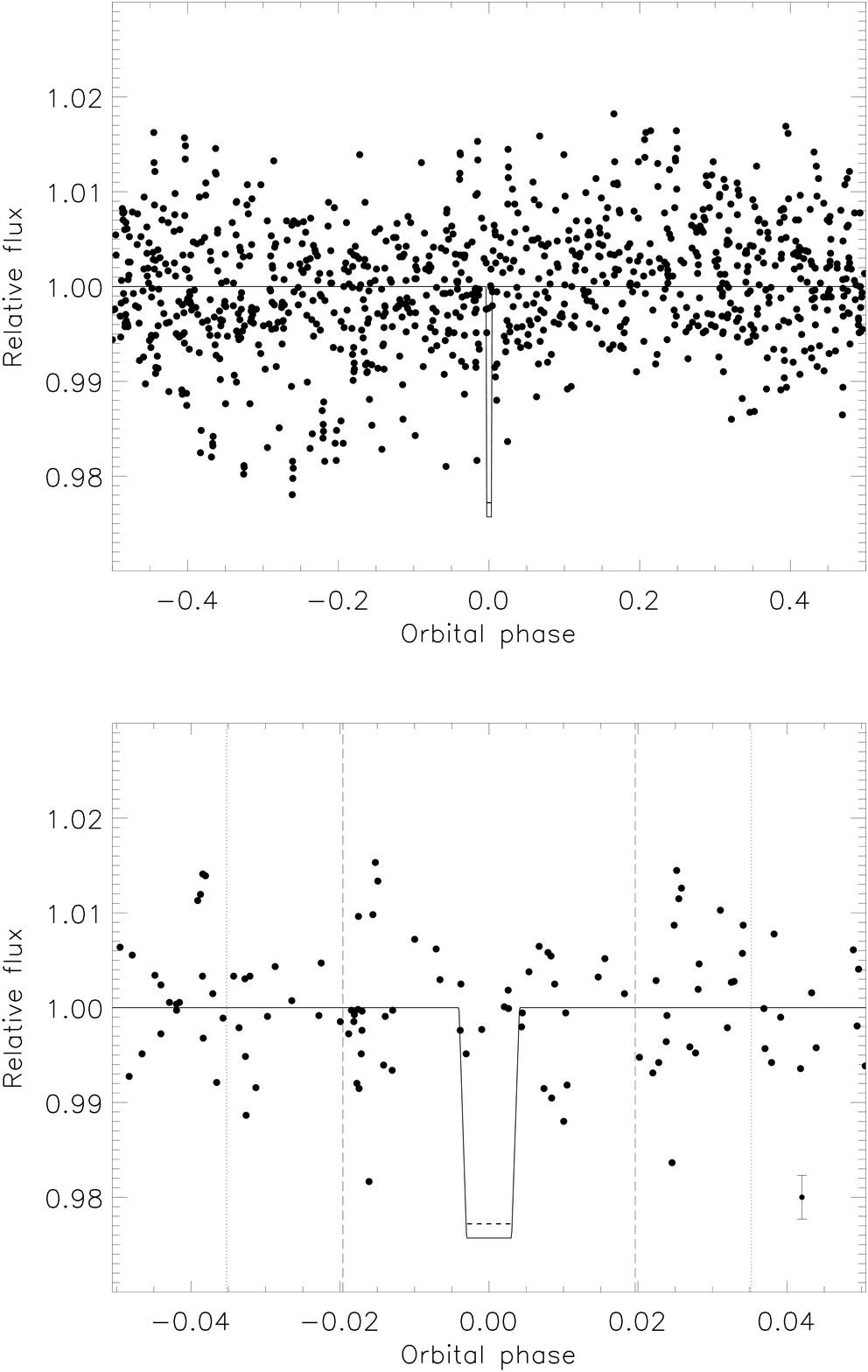}
\caption{APT photometry, with each half-season normalized to the same mean (see text for details) and phased at the orbital period of the planet. The solid line is the predicted transit signature. The overplotted short dashed line is the predicted transit signature if the star has a radius of 0.77 R$_{\odot}$. Top panel: full orbital phase. Bottom panel: the 3$\sigma$ transit window (horizontal extent of the plot), with the vertical dashed and dotted lines enclosing the 1$\sigma$ and 2$\sigma$ transit windows, respectively. The 1$\sigma$ error bar is shown in each panel. See section 6 in the text for details.}
\end{flushleft}
\end{figure}


\section{Conclusions}

The planet orbiting HD 192263 is an example of the challenges posed by the detection of companions around active stars. In this paper, we presented five new RV measurements which we used to refine the orbital parameters of HD 192263b. The new measurements provide continuing support for the existence of the planet, in agreement with the conclusion of \cite{San03}. 

We have also shown new photometry of the system. We perform a Fourier analysis of the photometry and find evidence for variability with periods near ~23 days, which agrees with previous reports of stellar variability and which we attribute to stellar rotation. A detailed examination of the Fourier spectrum near this value reveals a multiplet of peaks with periods ranging between 22 and 27 days. The dominant signal in this cluster has a period of 23.3932 $\pm$ 0.0046 days. The identified surrounding peaks arise from the evolving nature of spots on the stellar surface, which affect the shape and amplitude of the light curve on the timescale of the stellar rotation period. Nevertheless, neither the dominant period nor any of the five other peaks match the orbital period of the planet ($P=24.3587 \pm  0.0022$ days), so we find no evidence of star-planet interactions. 

We also observe a longer-term trend which may be a $\sim$ 8-year activity cycle (if the cycle repeats), or may just be part of a longer interval of random variations in the star's spottedness. Continuing long-term monitoring of the star should more convincingly discriminate between the two scenarios. It is noteworthy that we do not see a long-term variation in the RV measurements in phase with this long-term photometric trend. 

As our photometric dataset spans approximately a decade, we have good coverage of the 3$\sigma$ transit window when the data are phased to the orbital period of the planet. Thus we are able to thoroughly exclude transits of the predicted depth ($2.53 \%$, corresponding to a planet with a radius of 1.09 $R_{J}$) for a planet with a mass of 0.733 $M_{J}$. The absence of a detectable ($> 12.0$ m s$^{-1}$) Rossiter-McLaughlin effect is consistent with this result. We also exclude transit depths as low as $1.3\%$ (corresponding to a planetary radius of 0.79 $R_{J}$). In the case of a non-edge-on orbital configuration, the cadence of the data allows us to rule out transits with impact parameter $<$ 0.86.


\section*{Acknowledgements}

We thank Victoria Antoci for insightful conversations on interpreting the stellar variability and the anonymous referee for suggestions which have significantly improved the paper. The authors acknowledge financial support from the National Science Foundation through grant AST-1109662. D.D. is supported by a University of British Columbia Four Year Fellowship. The Center for Exoplanets and Habitable Worlds is supported by the Pennsylvania State University, the Eberly College of Science, and the Pennsylvania Space Grant Consortium. G. W. H. acknowledges support from NASA, NSF, Tennessee State University, and the State of Tennessee through its Centers of Excellence program.


\end{document}